\documentclass[12pt,letterpaper]{JHEP}
\usepackage{epsfig}
\newfont{\frak}{eufm10 scaled 1200}

\newfont{\Bbb}{msbm10 scaled 1200}     %instead of eusb10
\newcommand{\mathbb}[1]{\mbox{\Bbb #1}}
\DeclareSymbolFont{AMSa}{U}{msa}{m}{n}
\DeclareSymbolFont{AMSb}{U}{msb}{m}{n}
\let\Box\relax
\DeclareMathSymbol{\Box}{\mathord}{AMSa}{"03}
\def\IZ{{\mathbb Z}}
\def\IR{{\mathbb R}}

\def \da9{D9-\bar{D}9}
\def \dap{Dp-\bar{D}p}
\def \a{\bar{D}}

\def \cm{{\cal{M}}}

\def \n{\noindent}
\title{Cosmological Creation of D-branes and anti-D-branes}
\author{Mahbub Majumdar, Anne-Christine Davis\\
  DAMTP, Centre for Mathematical Sciences\\
   University of Cambridge, Wilberforce Road, Cambridge,  CB3 0WA, U.K.\\
  E-mail: \email{M.Majumdar@damtp.cam.ac.uk, A.C.Davis@damtp.cam.ac.uk}}

\abstract{We argue that the early universe may be described by an
initial state of space-filling branes and anti-branes.  At high
temperature this system is stable.  At low temperature tachyons appear
and lead to a phase transition, dynamics, and the creation of
D-branes. These branes are cosmologically produced in a generic
fashion by the Kibble mechanism.  From an entropic point of view, the
formation of lower dimensional branes is preferred and $D3$
brane-worlds are exponentially more likely to form than higher
dimensional branes.  Virtually any brane configuration can be created
from such phase transitions by adjusting the tachyon profile.  A lower
bound on the number defects produced is: one D-brane per Hubble volume.}

\date{\today}
\keywords{cosmology, branes, tachyon}

\preprint{DAMTP-2002-12}
\begin{document}

%%%%%%%%%%%%%%%%%%%%%%%%%%%%%%%%%%%%%%%%%%%%%%%%%%%%%%%%%%%%%%%%%%%%%%%%%%%%
%          Table of contents automatic !!!                                 %
%%%%%%%%%%%%%%%%%%%%%%%%%%%%%%%%%%%%%%%%%%%%%%%%%%%%%%%%%%%%%%%%%%%%%%%%%%%%
\section{{\bf Introduction}}

In this article we introduce the physics of brane anti-brane systems
to cosmologists and their cosmology to string theorists.  Recently,
there has been much interest in the cosmology of branes.  However,
little has been written about their origin and their anti-particles
(anti-branes) which are literally the flip side of branes.

Branes and anti-branes are unlike the point particles of the standard
cosmology; they are inherently non-perturbative.  They are solitons
and are very heavy at weak coupling.  If a strongly coupled phase
existed during the very early universe, branes may have been
perturbative objects in the coupling $1/g_s$.  When $g_s \rightarrow
\infty$, then $m_{Dp}\sim 1/g_s \rightarrow 0$ and branes become
virtually massless and may have been pair produced\cite{riotto}.
Because of their intricate substructure (they possess a gauge theory
on their world-volume), their production may have been entropically
more favorable than the production of elementary particles.  Large
string coupling corresponds to a large eleventh dimension since,
$g_s \sim R_{11}$, implying that a strongly coupled universe would likely
be eleven dimensional and described by M-theory ~\cite{wittenmtheory}.
A universe originating from a M-theory initial state would thus
presumably be populated by large numbers of branes of different
dimensionalities.

However, little is known about such a strongly coupled era and M
theory. Thus, it is difficult to describe the pair production of
branes and anti-branes in any concrete way.  But, non-perturbative
objects such as monopoles and cosmic strings are thought to have been
produced in the early universe through well understood processes -- as
topological defects created during various phase
transitions~\cite{shellard}.  In this paper we show that branes and
anti-branes may have been easily produced in a similar way -- through
a tachyonic phase transition, obviating the need for pair production
via a Schwinger type mechanism.  This so-called tachyonic phase
transition is well known, and has so far been thought of as a formal
mathematical device.  We will elevate it to a dynamical mechanism
motivated by certain ideas from topological K-theory.

A universe in which branes are produced should also contain
anti-branes.  A brane differs from an anti-brane only by possessing an
opposite (Ramond-Ramond) charge.  The charge of a D-brane corresponds
to an orientation.  Hence, a $\bar{D}p$ anti-brane is simply a brane
flipped over (rotated by $\pi$)~\cite{polchinski}.  It is hard to
understand why the early universe may have preferred a particular
orientation.  Thus, if branes were ubiquitous early on, it is likely
that anti-branes were too.  From a technical standpoint, anti-branes
are also required if the early universe was compact and cosmologically
unusual objects such as orientifolds were absent. Tadpole cancellation
requires them~\cite{srednicki}. (Orientifolds are unusual because they
represent boundaries of the universe and lead to non-local effects.)

Phase transitions are ideal for producing topological defects.  But
they often lead to an embarrassment of riches: too many are produced.
For example, in many GUT phase transitions enough monopoles are
created to dominate the energy density of the universe and cause it to
re-collapse~\cite{kolb}.  We find that generic tachyonic phase
transitions producing branes lead to similar problems: over-abundances
of branes and anti-branes.

In this paper we will try to answer the following questions: (1) were
D-branes produced after the big bang; (2) if so, how; (3) do they lead
to a brane problem, analogous to the monopole problem causing the
universe to re-collapse?

Most of our discussion refers to a universe modeled by Type IIB string
theory.  Generalizations to Type IIA or Type I theory are
straightforward.  The picture we construct of brane anti-brane
cosmology is the following.

$\bullet$ The universe starts as a stack of space-filling $\da9$ brane
anti-brane pairs.  The $\da9$ pairs may either be thought of as
describing the initial state of the universe, or as produced during a
high temperature era when large dimensional branes were most easily
created.

$\bullet$ At high temperature, $T>T_{H} /\sqrt{g_s N}$, the system is
stable~\cite{danielsson}.  No tachyons, which are present in brane
anti-brane systems at zero temperature, exist.  ($T_H$ is the Hagedorn
temperature, $N$ is the number of initial coincident $\da9$ brane
anti-brane pairs, and $g_s$ is the closed string coupling).  Because
of the brane tensions, and matter on the branes (gas of open string),
the branes expand and the temperature decreases. As the temperature
drops the potential of open strings transforming under the
bi-fundamental representation of the $U(N) \times U(N)$ gauge symmetry
changes.

$\bullet$ At $T \sim T_{H} / \sqrt{g_s N}$ the potential for those open
strings develops a double well structure and the strings become
tachyonic.  The tachyons then condense by rolling to the bottom of
their potential where the $U(N) \times U(N)$ gauge symmetry is broken
to a smaller group $G$.

$\bullet$ Upon tachyon condensation, the $D9$ branes and $\bar{D}9$
branes disappear. The tachyons cause their annihilation~\cite{sen}.
Cosmological topological obstructions via the Kibble mechanism
generically arise.  Once the tachyons roll down to the minima of their
potential, $D(9-2k)$ branes corresponding to non-zero homotopy groups,
$\pi_{2k-1}({\cal{M}}=U(N))$ are produced as topological defects.
Exotic spatial profiles for the tachyon field at the bottom of its
potential, $V(t)$, may lead to exotic brane configurations, such as
intersecting branes, etc.  Because the universe at this stage is
weakly coupled and at a lower temperature, entropy arguments favor the
formation of lower dimensional defects over higher dimensional
ones. In particular $D3$ braneworlds are exponentially more likely to
form than higher dimensional branes.  In Type IIA theory, $D(2p)$
branes are generically produced by the Kibble mechanism.  Unstable
non-BPS $D(2p)$ ($D(2p+1)$) branes in Type IIB (Type IIA) may also be
formed for $p$ an integer.  However, in the absence of any
stabilization mechanisms (orientifolds, etc.), they soon decay within
a few string times after their formation.

$\bullet$ The production of branes occurs cosmologically and in a
generic fashion.  By causality, at least one brane per Hubble volume
is created.  These branes lead to a brane problem and severe
cosmological difficulties.

String theory is not well understood in curved spaces.  Thus we have
made efforts to insure that most of our results are independent of the
spacetime metric and are model independent. For example, tachyon
condensation is a background independent process, as the shape of
tachyon potential is fixed.  Only a prefactor of the potential, the
brane tension, is model dependent. The formation of defects is a
topological process depending on the homotopy groups of the tachyon
potential. Entropy arguments based on bulk properties of branes are
used to postulate that lower dimensional branes are favored as end
products of tachyon condensation.  Also, the number of $Dp$ branes
filling spacetime -- one per Hubble volume -- is an argument based on
topology and causality.

The plan of the paper is as follows.  In section \ref{initial} we
examine whether the universe, modeled by Type IIB theory, may have
begun from an initial state of $\da9$ brane anti-brane pairs. In
section \ref{properties} we describe the properties of a $\da9$
system, and more generally the properties of brane anti-brane
systems. Next, in section \ref{production} we explain how lower
dimensional branes may have filled the early universe. In section
\ref{typeIIAtheory}, we quote the analogous results for Type IIA
theory.  Finally, in section \ref{conclusions} we address some
possible objections to our results.

\section{An Initial State of $N$ $\da9$ Brane pairs }\label{initial}

\subsection{Motivation} \label{motive}

In recent years there has been a great multitude of brane universe
models.  The original ADD scenario started out as a flat D-brane in
Minkowski space~\cite{dvali}. Then came the Randall-Sundrum models:
brane(s) embedded in adS~\cite{randall}.  Subsequently, many variations
appeared, like the manyfold universe (a brane folded on top of itself),
and intersecting brane models~\cite{others}.  Such models, though
apparently disparate, possess enough unifying characteristics to allow
a study of the whole gamut of models using only a few tools.  They can
all be built out of pairs of $D9$ branes and $\a9$ anti-branes in Type
IIB string theory, or the unstable $D9$ branes of Type IIA string
theory~\cite{minasian, witten, horava}.

Recently, the K-theory classification of D-brane charges has taught us
that any D-brane configuration in type IIB string theory can be built
out of $N$ $\da9$ brane anti-brane pairs.  On their world-volumes, the
9-branes and anti-branes possess tachyon fields and fluxes akin to
electromagnetic fluxes.  One starts with a collection of $\da9$ branes
and adjusts the tachyon fields and fluxes on the 9-branes.  A phase
transition then occurs turning the original constituents into the
configuration of choice.  D-brane configurations in Type IIA string theory
can similarly be constructed from unstable $D9$ branes and their
world-volume fields, although subtleties arise requiring the use of
K-homology~\cite{khomology}. Hence, we can study any brane world
scenario by studying 9-branes and the tachyons and fluxes on them.

Such 9-branes may be thought of as mathematical devices to create
brane configurations.  However, they can also be taken to be
real. We treat them as physical objects and model the early universe
as emerging from an initial state of $\da9$ branes.  (We pick Type IIB
theory as our starting point.)

The $\da9$ branes may be thought of as an initial state with the
following properties.  It is an open string vacuum state, since $\da9$
pairs are a solution to the ``equations of motion'' of open strings.
This vacuum is unstable at zero temperature.  But, it is thought to be
stable at high temperature. Since the branes are space-filling, there
is no ad hoc partition of spacetime into Neumann and Dirichlet
directions.  Hence, apart from compactification issues, such a state
describes an isotropic and homogeneous universe.  Also, such an open
string vacuum state necessarily contains open and closed string
excitations. Closed strings correspond to gravity and open strings
correspond to gauge fields like photons or gluons and one expects the
early universe to have possessed gravitational and gauge degrees of
freedom.  Gravity appears on the space-filling branes because of
worldsheet duality. This duality implies that open string theories
necessarily contain closed strings. In the case of $Dp$ branes, the
closed strings are in the bulk. However, for space-filling branes,
there is no bulk, or rather the bulk coincides with the world-volume
of the branes~\cite{joebigbook}.

Previous string cosmology scenarios such as the pre-big bang proposal,
have always taken the closed string vacuum state as an initial
state~\cite{veneziano}.  Closed string theories do not necessarily
contain open strings (gauge fields).  Thus the description of the
early universe as an open string vacuum seems more natural.

The rationality of an initial state of $N$ $\da9$ branes can also be
argued from thermodynamics.  It appears entropically
favorable to create large dimensional branes at very high energy
densities and {\em weak coupling}~\cite{kogan}.

\EPSFIGURE[l]{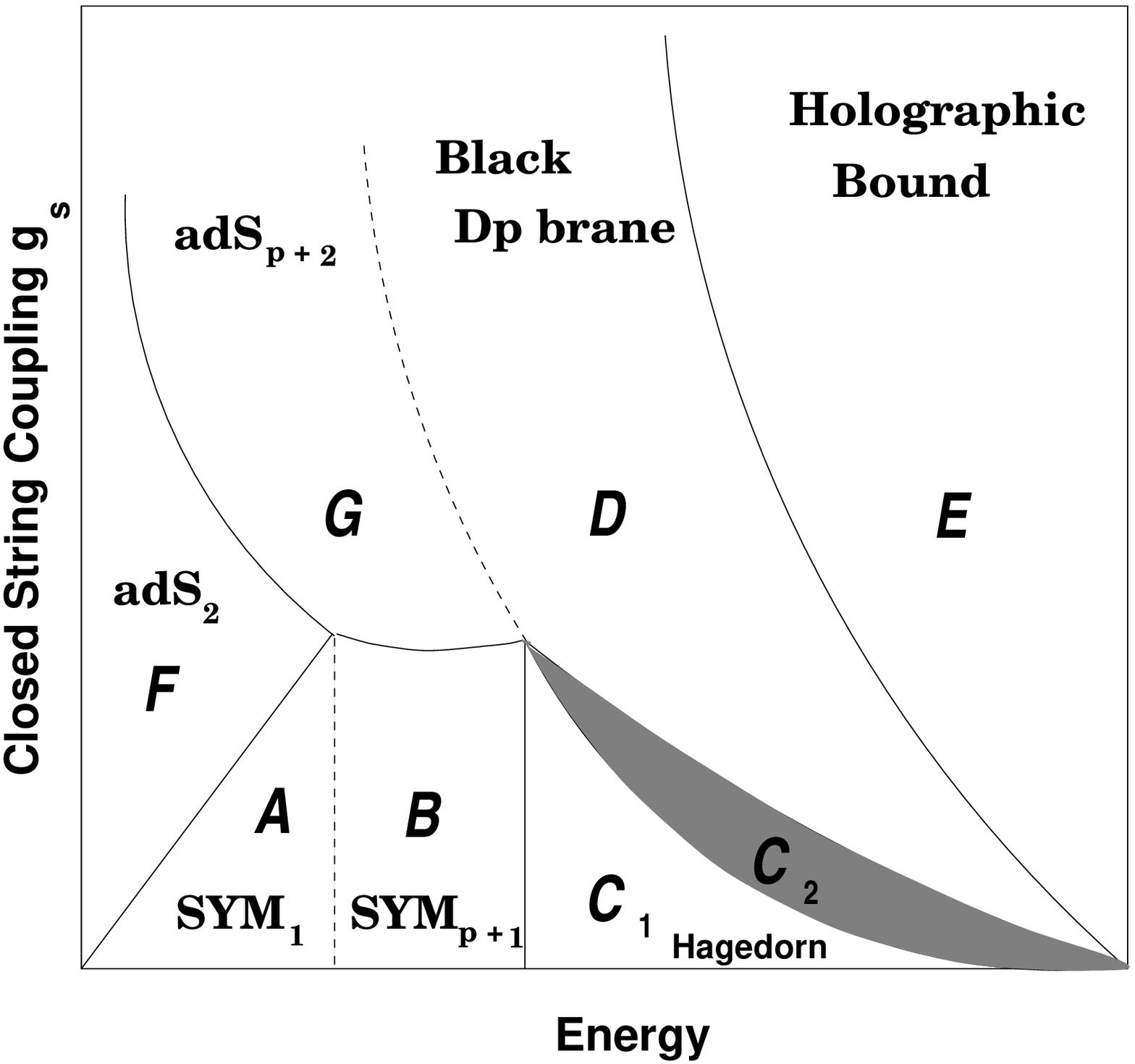,width=7cm}{The phase diagram of D-branes
as a function of the closed string coupling, $g_s$ and energy, $E$.
\label{figure1} }

We recall the D-brane phase diagram discussed in Abel, Barbon, Kogan
and Rabinovici ~\cite{kogan,kogan2,martinec}.  A D-brane admits two
different thermodynamic descriptions.  At weak coupling, the entropy
of a brane is dominated by the gas of open (super Yang-Mills) strings
on its world-volume, which corresponds to regions $A$ and $B$ in
figure \ref{figure1}.  If the temperature is raised, the increased
energy density is transferred to the open strings. Since the mass
of the strings is proportional to their length, the open strings grow
and start to explore the surrounding space.  If compact directions
exist, the strings may wind these directions leading to winding
modes.\footnote{These open string winding modes dominate the energy
density because their heat capacities diverge, and they can be fed
large amounts of energy without significantly raising their
temperature. They are limiting systems and cannot reach the Hagedorn
temperature.  Closed strings are non-limiting, and as $T\rightarrow
T_{H}$ all the energy flows into the open strings~\cite{frau}.}  This
corresponds to the Hagedorn regime, regions $C_1$ and $C_2$. At very
high energy density the energy of a D-brane is dominated by the cloud
of open strings on it. In region $C_2$ the energy of the gas equals,
and exceeds, the rest mass of the D-brane.  If the energy is raised
further, the Jeans mass is exceeded and a black hole forms, region
$D$.

At strong coupling D-branes have a different description. The entropy
of a brane in this regime is dominated by its geometrical entropy --
akin to the area of a black hole.  A finite temperature $Dp$ brane has
horizons. At low temperatures it has an $adS_{p+2}$ geometry, region
$G$.  At sufficiently low temperatures the brane can be described as a
collection of ``smeared'' $D0$ branes whose geometry is $adS_2$,
region $F$.  At very high temperatures, the horizons of the
$adS_{p+2}$ geometry shift and the $Dp$ brane turns into a black hole,
region $D$.  As the energy is further increased the black hole grows
until it fills up the entire space and reaches the holographic bound,
region $E$.

The dotted line partitioning regions $C_1$ and $C_2$ denotes when the
energy $E$ of the open string gas becomes larger than the $Dp$ brane
mass, $m_{Dp} \sim 1/g_s$.  This region may be unstable to the
spontaneous production of D-branes or anti-branes leading to the
screening of any already existing Ramond-Ramond charge.  Abel, et al.,
~\cite{kogan} assert that in thermal equilibrium the following reaction
should be common in region $C_2$: $D + \bar{D} \leftrightarrow X +X$,
where $X$ are massless fields.  The chemical potential of the massless
fields is zero implying that in equilibrium, $\mu + \bar{\mu} = 0$,
where $\mu$ and $\bar{\mu}$ are the chemical potentials of the branes
and anti-branes respectively.  In the absence of any CP
violation\footnote{We thank Phillipe Brax for mentioning this
caveat to us.}, and for significant brane production, symmetry dictates that
$\mu = \bar{\mu} = 0 $.  Because of their considerable structure (a
world-volume gauge theory), branes can be crudely modeled as particles
having internal structure and obeying Boltzmann statistics.  The
chemical potential is then

\begin{equation}
\mu  \sim \ln \left ( \frac{V_{\perp}}{N} \sum_{\vec{p}_{\perp}, n} e
^{-\beta E(\vec{p}_{\perp}, n)} \right )
\label{chempot}
\end{equation}

\n
The sum is over the center of mass motion of a brane and its quantum
numbers $n$. The energy is $E(\vec{p}_{\perp}, n)= M_p +
\vec{p}^2_{\perp}/2m_{Dp} + \epsilon_n$. The $\epsilon_n$ are the
internal excitations and represent open string excitations on a
D-brane.  The free energy of the open string gas on a $Dp$ brane is
$e^{-\beta F_g}= \sum_n e^{-\beta \epsilon_n}$.  If $\mu=0$,
eq. \ref{chempot} can be inverted giving,

\begin{equation}
\frac{V_{\perp}}{N} \sim e^{-\beta m_{Dp}} \sum_{n} e^{-\beta
\epsilon_n}\int d\vec{p}_{\perp} e^{ -\beta \vec{p}^2_{\perp}/2m_{Dp}}
= e^{-\beta (m_{Dp} + F_g)} \left ( \frac{m_{Dp}}{\beta} \right
)^{d_{\perp}/2}
\end{equation}

\n Near the Hagedorn temperature for $p>6$, the free energy, $F_g$,
diverges negatively. Thus unsuppressed pair production of large
dimensional branes may occur in Type IIB string theory.  The free
energy diverges because, above $p=6$ the system is {\em limiting}.
Increases in energy lead to more winding modes and an energy pile-up on
the branes. Note, the density of states is proportional to the (volume
of the brane)/(volume of the transverse space). For $p<6$, the system
is not {\em limiting}, and can lose its energy to the closed string
bath (in the bulk).

It would have been impossible to spontaneously produce heavy
defects like D-branes at high temperature, had they not possessed a
world-volume gauge theory.  Because of their intricate structure it is
entropically favorable to produce them at high temperature. We can
make this more obvious by focusing on the gauge group on the branes
and noting how it influences the free energy~\cite{danielsson}.

Consider a configuration of $N$ branes and $N$ anti-branes.  Let $k_1$
and $k_2$ be two positive numbers less than unity.  Suppose $k_1 N$
branes are coincident.  Then a $U(k_1 N )$ gauge theory will appear
increasing the entropy by a factor of $k_1 N $.  If $k_2 N $ brane
anti-brane pairs are coincident, then (tachyonic) open strings
transforming in a bifundamental representation $(k_2 N, \overline{k_2
N})$ will arise. This will increase the entropy by a factor of $k_2^2
N^2$.

Let $F_{Dp}$ be the free energy due to the center of mass motion of
branes and anti-branes, and $F_{photon}$ be the entropy of the open
string excitations on non-coincident branes and anti-branes.  Suppose
the number of branes which are coincident equals the number of
anti-branes which are coincident and let $F_{gluon}$ be the free
energy of a single pair of coincident branes. Let $F_{tachyon}$ be the
free energy due to open string excitations on a coincident brane
anti-brane pair.  Finally let $F_{V(t)}$ be the free energy due to the
potential energy of the tachyons.  Then the free energy of the open
string and D-brane system is

\begin{equation}
F = F_{V(t)} + F_{Dp} + F_{photon} + 2 k_1 N F_{gluon} + k_2^2 N^2
F_{tachyon}
\end{equation}

\n $F_{photon}, F_{gluon}$, and $F_{tachyon}$ are the free energies of
open string gases on the brane, and as discussed above, diverge
negatively near the Hagedorn temperature for large $p$.  Maximizing
the entropy means minimizing the free energy, and the free energy as a
function of $N$ looks like an inverted parabola.  It is minimized by
taking $N$ to infinity.  This would seem to indicate that, at least in
the canonical ensemble, where the system is connected to an infinite
reservoir of energy, that configurations with large numbers of high
dimension branes are probable.

\subsection{Supersymmetry}

A system of branes and anti-branes such as a pair of $\da9$ breaks all
supersymmetries ~\cite{sen}.  This may be disconcerting.  However,
from our point of view, it may actually be beneficial.  Supersymmetry
is valuable for solving the hierarchy problem at low energies.  But it
is problematic for cosmology because it is difficult to reconcile with
time dependent metrics~\cite{malda}.  For example, in the simplest set
up with no non-perturbative fluxes turned on, supersymmetry requires
the existence of a timelike Killing spinor.  A timelike Killing spinor
roughly squares to a timelike Killing vector.  Systems with timelike
Killing vectors are stationary, and are cosmologically not very
interesting.

Brane anti-brane systems are non-supersymmetric at high energies, but
after undergoing tachyonic phase transitions can easily flow to
supersymmetric configurations at lower energies.  By adjusting the scale at
which the transitions take place, one can adjust the energy at which
the supersymmetry needed for the hierarchy problem is restored.  Thus
by breaking supersymmetry at high energies, time dependent vacuum
solutions of the supergravity equations may be more easily generated
and the hierarchy problem resolved by restoring supersymmetry at lower
energies. The time dependence of the universe at late times would be a
separate problem to solve. But it would be no harder than it
is already.

\section{Properties of $\da9$ Systems}\label{properties}

\subsection{The Presence of Tachyons at Zero Temperature}

\EPSFIGURE[l]{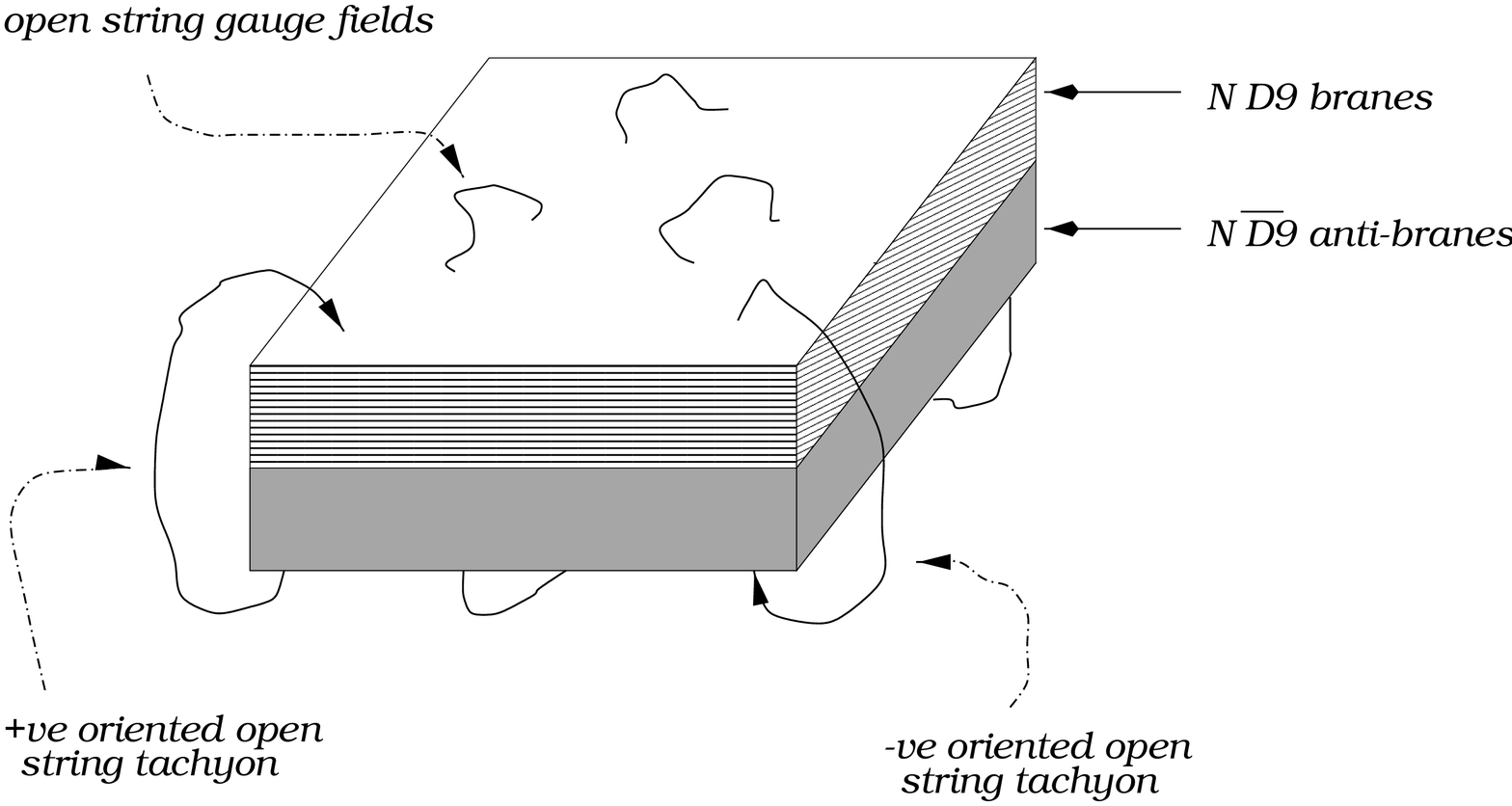,width=8cm}{$N$ coincident $\da9$
brane anti-brane pairs; open strings stretching between branes and
anti-branes are tachyons.\label{braneconfig}}

Brane anti-brane systems are tachyonic at zero
temperature~\cite{susskind}.  This is because the GSO projection on the
sector of open strings connecting the branes to the anti-branes is
reversed~\cite{sen}.  Thus, instead of projecting out the tachyons in
this sector, the GSO projection insures that the tachyons remain. In
fact, there are two tachyons per brane and anti-brane, because strings
connecting the branes to anti-branes can be oriented in two ways, see
figure \ref{braneconfig}. Each pair of tachyons can be complexified
into a complex tachyon field.  For $N$ coincident branes and
anti-branes, there are $N^2$ ways of connecting the $N$ branes to the
$N$ anti-branes. Thus, the tachyons transform in the bifundamental
representation $(N, \bar{N})$ of the gauge group on the brane: $U(N)
\times U(N)$.

The existence of tachyons may be surprising as superstring theory
claims to be tachyon-free.  However, these tachyons are manifestations of
the instability one expects from particle anti-particle
attraction.  The tachyon eventually leads to the annihilation of the
brane and anti-brane, provided no topological obstructions to
annihilation exist~\cite{sen}.

The tachyon field, $t$, possesses a potential $V(t)$ which depends on
only $|t|$.  The potential has a double well shape.  Initially, the
tachyon starts at the top of the potential, $t=0$, and then rolls down
to the bottom, $t_0$. If no topological obstructions to tachyon
condensation exist, then at the bottom of the well, $t= t_0$, the
negative tachyon energy cancels the tensions of the brane and
anti-brane leaving the manifestly supersymmetric closed string vacuum.
Specifically, for $N$ coincident brane anti-brane pairs, as
Sen~\cite{sen} has argued and has been verified by ~\cite{sen2,
gerasimov,sigma}:

\begin{equation}
V(t_0) + 2N\tau_p = 0.
\end{equation}

When $t=t_0$ the phase transition ends. Since the transition is second
order, all of spacetime flows from the open string vacuum to the
closed string vacuum state in a finite time.  No physical open string
degrees of freedom will be left and none of the false vacuum (open
string vacuum) remains. The open string degrees of freedom represented
by the gauge fields on the brane acquire masses and become confined.
Closed string dilatonic/gravitational/RR-radiation fills the space.

A vital feature of the tachyon potential is that it can universally
be written as ~\cite{sen3}

\begin{equation}
V(t) = 2\tau_p(1+v(t)).
\label{uni}
\end{equation}

\n Here $v(t)$ is a dimensionless universal function of the tachyon
field.  It is exactly -1 at the bottom of the potential.  For
superstrings, the bottom is always a global minimum.  Apart from the
multiplicative factor, $\tau_p$, the tachyon potential is the same for
flat D-branes, D-branes wrapped on cycles of an internal compact
manifold, D-branes in the presence of a background metric or
anti-symmetric tensor field. The tension, $\tau_p$, is the only model
dependent parameter and depends on the boundary conformal/string field
theory used to describe the system.

\subsection{Remarks on Open String Tachyons at High Temperature}

\EPSFIGURE[l]{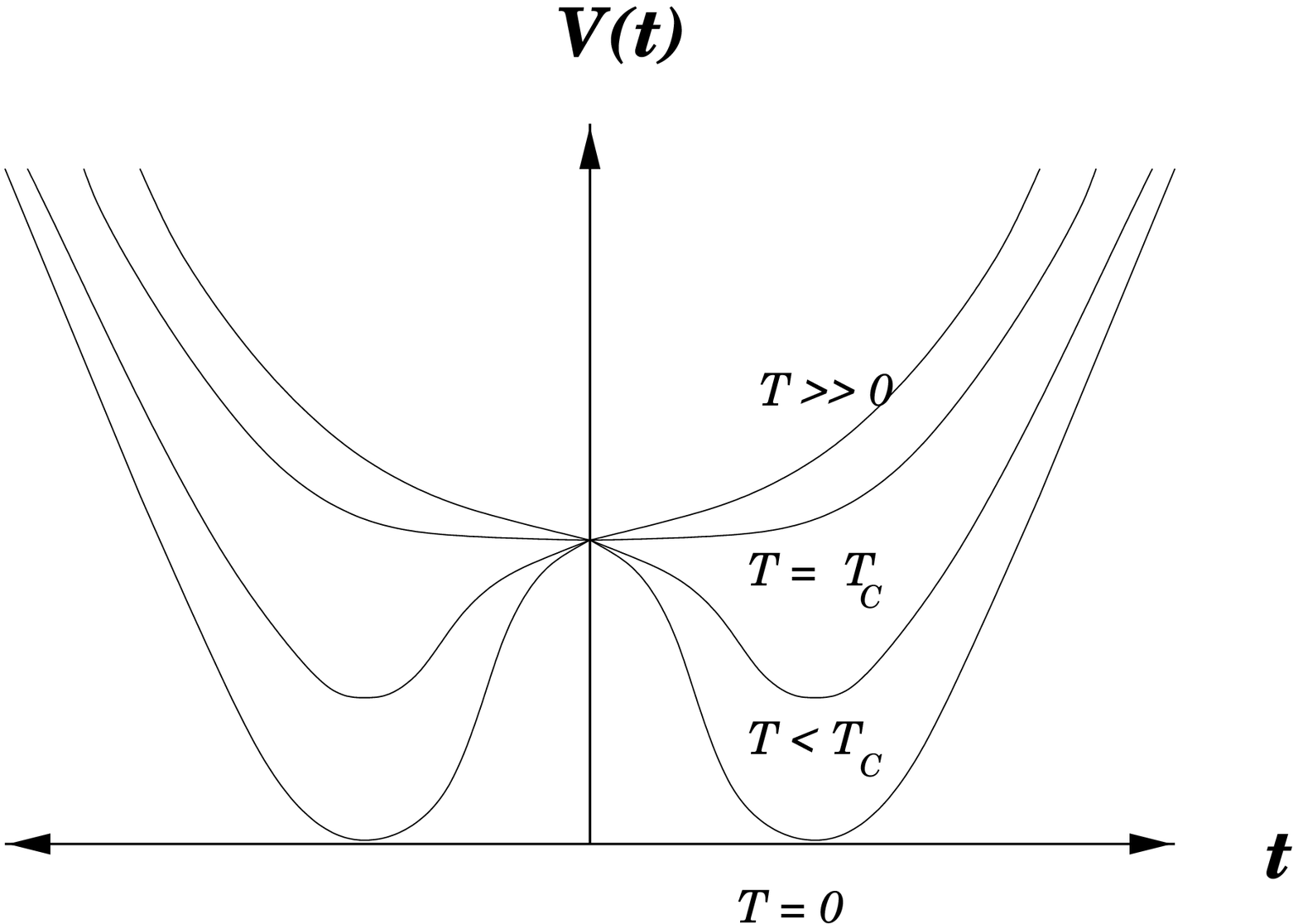,width=7cm}{The tachyon potential
$V(t)$ above and below the critical temperature $T_c$ $\sim
T_H/\sqrt{g_s N}$. \label{thermalpotential}}

The behavior of the tachyon potential at high temperature is not well
understood.  However, we make some remarks based on the preliminary
investigation by Danielsson et al. ~\cite{danielsson}.

In field theory, it is well known that finite temperature loop
corrections lead to symmetry restoration at high
temperature~\cite{kapusta}.  For example a double well potential for a
scalar field $\phi$ in a $\phi^4$ type theory will slowly change shape
and become parabolic at some critical temperature. See figure
\ref{thermalpotential}.

The authors argue in ~\cite{danielsson} that something similar happens
to the tachyon potential.  This is very significant for $\da9$
physics. In a parabolic potential, the tachyon is no longer
``tachyonic'' and possesses a positive mass, implying that brane
anti-brane systems are stable at high temperature.  This insures that
a universe composed of a stack of $\da9$ branes is initially stable.
The physical reason cited for stability is the following: by shifting
the equilibrium value of the tachyon field upwards toward the
symmetric point, the tachyon mass decreases. Although, it costs energy
to do this, a gas of smaller mass tachyons has a larger entropy.  At
very high temperature, the minimum of the potential can shift to
the symmetric point, $t = 0$.

More concretely, the symmetric point minimizes the free energy at
temperatures above a critical temperature $T_c$

\begin{equation}
T\ge  T_c \sim \frac{T_H}{\sqrt{g_sN}}
\label{tc}
\end{equation}
\n
where $g_s$ is the closed string coupling and $T_H$ is the Hagedorn
temperature.  This can be shown by considering the potential and
tachyon parts of the free energy, $F_{V(t)}$ and $F_{tachyon}$
respectively.  For simplicity, consider the case of a $D3-\bar{D}3$
pair, and let the tachyon potential be parameterized as in
eq. (\ref{uni}).  Then,

\begin{equation}
F = 2 \tau_3 {\rm Tr} v(t)  + \frac{\Omega_2}{(2 \pi)^3} c N^2 \beta^{-4}
\int_0^{\infty} dxx^2 \ln \left ( \frac{1- e^{-\sqrt{x^2 + \beta^2 m^2}}}{1+ e^{-\sqrt{x^2 + \beta^2 m^2}}}
\right )
\label{freeenergy}
\end{equation}

\n where $\tau_3$ is the 3-brane tension and $m$ is the mass given to
the gauge bosons which gain a mass and become confined due to tachyon
condensation. Here $c=8$, and is a degeneracy factor accounting for
the 8 bosonic and 8 fermionic degrees of freedom. By starting
at the symmetric point and letting one of the diagonal tachyons
condense by an amount $\delta t$, the tachyon gains mass, the gauge
bosons get a mass $\delta m$, and the free energy changes
(\ref{freeenergy}) by

\begin{equation}
\delta F = -4 \tau_3 (\delta t)^2 + \frac{1}{2} \frac{N}{\beta^2}
(\delta m)^2
\end{equation}

\n For sufficiently large temperature, $\delta F$ is no longer
negative for a tachyon rolling towards its minimum $t_0$.  It becomes
positive for $T > T_c$.  Thus, if tachyon condensation occurs above
$T_c$, the entropy decreases.  Therefore, tachyon condensation
does not occur, and the minimum of the tachyon potential can shift
from $t=t_0$ to $t=0$.

However, this high temperature description of the open string vacuum
($t=0$) is typically non-perturbative. The Hagedorn temperature for
open string systems is {\em limiting} and the temperature on the brane
cannot exceed the Hagedorn temperature, $T_H$.\footnote{In some cases
the phase transition can occur at the Hagedorn temperature. For
example, Kogan has mentioned to us that in some realistic $SU(5)$
models at grand unification we can have $g_s N \sim 1$ with $g_s<1$ at
the transition.  Stringy Hagedorn inflation then occurs near the
transition~\cite{hagedorninflation}.} Therefore, in order to satisfy
eq. (\ref{tc}), the 'tHooft coupling $g_s N$ must exceed
unity. Although the closed string coupling $g_s$ and the open string
coupling $g_s^{1/2}$ may be small, the effective ('tHooft) coupling,
$g_s N$, must be large. An exception to this has been suggested by
Kogan.

\subsection{Consequences of Topological Obstructions on Tachyon
Condensation}

If topological obstructions exist then the end product of tachyon
condensation will not be pure vacuum; lower dimensional branes will be
left over.  This picture is familiar from years of work on monopoles
and cosmic strings.  For example, in the case of the Abelian Higgs
Model, if the Higgs potential has non-zero winding such that the
fundamental group is non-zero, $\pi_1(U(1)) \neq 0$, then a cosmic
string will be left after the Higgs field rolls down to a minimum of
its potential~\cite{shellard}.  Similar things happen in the tachyon
case. Since the tachyon is a complex field, it can wind around a
co-dimension two locus and non-zero winding around the tachyon
potential leads to left over defects.  The type of D-brane which is
left over from tachyon condensation depends on which homotopy groups
are non-zero.

A vortex, the higher dimensional analog of a cosmic string, will be
left if $\pi_1(\cm)\neq 0$, where $\cm$ is the vacuum manifold of the
tachyon potential.  If the tachyons condense on $Dp$ and
$\bar{D}p$ branes, then a vortex is a $D(p-2)$ brane. Similarly, if
$\pi_3(\cm) \neq 0$, then a vortex of a vortex -- which is $D(p-4)$
brane will be left over.  More generally, if the tachyons of a system
of $\da9$ branes condense, a $9-2k$ dimensional D-brane will be
created if $\pi_{2k-1}({\cal{M}}) \neq 0$.

The symmetry group of the vacuum manifold $\cm$ is $U(N)$ if $N$
$\da9$ branes condense. This is because, the gauge symmetry is broken
to the diagonal subgroup $(U(N) \times U(N))/U(N) \sim U(N)$ once the
tachyons receive expectation values.  The existence of stable left
over defects from tachyon condensation can be inferred from the
presence of some Ramond-Ramond charge, which is measured by the
K-theory group, ${{\rm K}}^0$. For example, ${\rm K}^0$ tells us that
if the tachyons condense in $2k$ ($k\in \IZ^+$) flat dimensions, such
that false vacuum is left in $9-2k$ directions, that $9-2k$ D-brane
charge may be left since ${\rm K}^0(S^{2k}) = \IZ$.  Here, $S^{2k}$ is
the one point compactification of $\IR^{2k}$. Now, ${{\rm K}}^0$ can
be directly related to the vacuum manifold $U(N)$ through the homotopy
groups $\pi_{2k-1}(U(N))$ and used to tell us when tachyon
condensation onto $\cm = U(N)$ will leave $Dp$ brane
charge. Specifically~\cite{kgroups},

\begin{equation}
{{\rm K}}^0(S^{2k+r}) = \left\{
\begin{array}{cll}
\IZ& {\rm if} &r=0\\
0  & {\rm if} &r=1
\end{array}\right\} = \pi_{2k-1+r}(U(N))\ \ \ \
\forall N \ge k +r,~ k \in \IZ^+.\
\label{homotopygroup}
\end{equation}
\n

This means that if the tachyon field condenses in even dimensions
($r=0$) that stable $9-2k$ branes may form if $N \ge k$, and that if
the tachyon condenses in odd dimensions ($r=1$) that no stable branes
will form as long as $N \ge k+1$.  In fact no {\em stable} defects
will form for any $N$, if the tachyon condenses in odd dimensions. The
homotopy groups $\pi_{2k}(U(N))$ will not generally be trivial for $N
< k+1 $.  For example~\cite{krevaire}, if $N=k-1$ and $k \in 2 \IZ$,
then $\pi_{2k} (U(N =k-1)) = \IZ_{(2k!)/2}$. However, such a
$9-(2k+1)$-brane described by $\pi_{2k}$, does not possess any
Ramond-Ramond charge, as can be verified from the Chern-Simons part of
the action.  It will decay to vacuum in much the same way that the
global strings of cosmic string theory, which appear due to a
non-trivial $\pi_0$, decay via the emission of
radiation.\footnote{Note, even if it is physically possible to form
such branes, they are unlike the more familiar unstable branes of Type
IIB a la Sen, because they cannot decay to lower dimensional stable
branes as this would increase the co-dimension $2k+1$ while keeping
$N$ fixed.} Likewise, if $N < k$ and the tachyon condenses in even
dimensions, no physical/stable $D(9-2k)$ branes can form.  (A lower
dimensional $D(9-2k')$ brane can of course form, for $N \ge k'$.)
Thus by restricting to the {\em stable range} $N\ge k+r$, we obtain
all the physically possible/stable D-branes. Table \ref{homotopytable}
makes (\ref{homotopygroup}) more explicit.  This is a hint of the
power of K-theory.  It identifies all of the possible decay products
almost for free, whereas homotopy theory mistakes some
unstable/unphysical configurations as stable.\footnote{ If $N=2^{k-1}$
then a $D(9-2k)$ brane can be explicitly constructed as a vortex
solution. $N=2^{k-1}$ implies that $2^{k-1}$ $D9$ branes and $2^{k-1}$
$\bar{D}9$ anti-branes are used to produce a $9-2k$ dimensional
D-brane~\cite{witten}.  The role of the $2^{k-1}$ $\da9$ pairs can be
seen from the example of $2^2$ $D9-\bar{D}9$ condensing to a $D3$
brane. A single $\da9$ pair with winding around the tachyon potential
can always produce a $D7$ brane as a vortex.  If the tachyon
potentials of two $\da9$ pairs have positive winding, then each pair
will create a $D7$. If the potentials of the other two $\da9$ branes
have negative winding, then each pair will create a $\bar{D}7$. Hence,
two $D7-\bar{D}7$ pairs arise.  Suppose one such pair creates a $D5$
via positive winding around its tachyon potential, and the other pair
produces a $\bar{D}5$.  Then one $D5-\bar{D}5$ pair results, which can
produce a $D3$ brane-world as a vortex.  This stepwise construction
gives the same result as simply evaluating $\pi_5(U(4))$ for the four
$\da9$ pairs.}

\begin{table}
\begin{center}
\begin{tabular}{|c|c|c|c|c|c|c|c|c|c|c|} \hline
\ $Dp$-brane\ &  \ $D8$\ & \ $D7$\ & \ $D6$\ & \ $D5$\ & \ $D4$\ & \ $D3$\
& \ $D2$\ & \ $D1$\ & \ $D0$\ & \ $D(-1)$\ \\ \hline
Transverse Space &
$S^1$ & $S^2$ & $S^3$ & $S^4$ & $S^5$ & $S^6$ & $S^7$ & $S^8$
& $S^9$ & $S^{10}$ \\ \hline
{Homotopy Group} & $\pi_0$ & $\pi_1$ & $\pi_2$ & $\pi_3$ &
$\pi_4$ & $\pi_5$ & $\pi_6$ & $\pi_7$ & $\pi_8$ & $\pi_9$ \\ \hline
$K^0(S^n)= \pi_{n-1}(U(N))$ & 0 &$\IZ$ & 0
& $\IZ$ & 0 & $\IZ$ & 0 & $\IZ$ & 0 & $\IZ$ \\ \hline
Minimum $N$  & 1 & 1 & 2
& 2 & 3 & 3 & 4 & 4 & 5 & 5 \\ \hline
\end{tabular}
\end{center}
\caption{\em Type IIB D brane spectrum; the second line is the 1-point
compactified transverse space; minimum $N$ is the minimum number of
$\da9$ pairs needed to create a $Dp$ brane and is the beginning of the
stable range.}
\label{homotopytable}
\end{table}

At the beginning of section \ref{motive} we advertised brane
anti-brane systems as a way of creating any brane-world model by
adjusting fields on the brane anti-brane system's worldvolume.  Those
fields are the fluxes and the tachyons on a $\dap$ system.  Their
values determine whether the homotopy groups $\pi_{2k-1}(\cm) \neq 0$
and which branes form at the end of tachyon condensation.  The field
strength, $F$, in some sense provides the energy for the defects to
form.  The flux, and its generalizations are quantized. Each quanta
corresponds to a single brane.  For example, if $\int F$ over a 2
surface is unity, then $\pi_1(\cm)=1$ and a $D(p-2)$ brane results.
The tachyon can be thought of as determining the shape of the final
brane configuration.  Regions where the tachyon field is stuck at
$t=0$ (the false vacuum) become part of the daughter brane.  Regions
where the tachyon has condensed to $t=t_0$ become the bulk and no
brane exists there.  For example, if the tachyon vanishes on a
co-dimension two surface and has the value $t=t_0$ in the two
transverse directions, a vortex will form.  Generally, to produce a
certain configuration such as a $D3$ intersecting a $D5$, the tachyon
field must be forced to be zero on the configuration and have the
value $t_0$ in the transverse space.  One can metaphorically describe
the field strength as providing the ``paper'' out of which to create
the desired brane, and the tachyon as the scissors which allow one to
cut out the desired shape.

We can demonstrate this with the following example ~\cite{sen}: the
construction of a $D1$ brane as a vortex on a $D3-\bar{D}3$ pair.  An
effective Lagrangian for the tachyon on a $D3-\bar{D}3$ can be written
as ${\cal{L}} = (D_{\mu} t)^2 - V(t)$, where the gauge covariant
derivative is $D_{\mu} = \partial_{\mu} - i(A_{\mu} - A'_{\mu})$.
Here $A_{\mu}$ is the gauge field on the brane and $A'_{\mu}$ is the
gauge field on the anti-brane. A finite energy vortex will be realized
if at large $r$ the tachyon takes the value $t=t_0$ such that
$V(t)\rightarrow V(t_0)$, and the kinetic energy vanishes: $(D_{\mu} t
)^2 \rightarrow 0$.  The kinetic energy at large $r$ can be seen to
vanish if there is a net quantized magnetic flux on the $D3-\bar{D}3$,

\begin{equation}
\frac{1}{2 \pi} \int (F - F') d^3x = 1.
\label{fluxquant}
\end{equation}

\n Eq. \ref{fluxquant} is the tachyon field's winding number.  It is
also the first Chern class of the gauge field configuration. If
$A_{\mu}$ has cylindrical symmetry then eq. (\ref{fluxquant}) implies
that in the radial gauge $A_r=A'_r =0$ for large $r$

\begin{equation}
\oint_{{\rm large}\,\, S^1} (A_{\theta} - A'_{\theta}) d\theta = 2 \pi
\;\;\;\Rightarrow \;\;\; A_{\theta} - A'_{\theta}, \rightarrow 1
\label{gaugefld}
\end{equation}

Eq.(\ref{gaugefld}) and the fact that the tachyon takes values on the
vacuum manifold, $t \rightarrow t_ 0 e^{i\theta}$, imply that $(Dt)^2
\rightarrow 0$, and that a vortex solution exists.  Using the
Chern-Simons part of the action of the $D3-\bar{D}3$ pair, one can
show that the vortex has Ramond-Ramond charge and is a $D1$
brane~\cite{kennedy}.  Note, this construction is virtually the same as
the cosmic string solution of the Abelian Higgs model.

If one is interested in only counting the number of daughter branes
arising from tachyon condensation, one can map the K-theory group
classifying the D-brane charges into cohomology, and use the total
Chern class to find explicit expressions for embedded charges on a
$\dap$ brane configuration.  The total Chern class can be decomposed
into Chern classes, $c_i$, whose integrals measure the fluxes and
their generalizations on a brane anti-brane system.  For example, the
Chern classes of the system are defined as

\begin{eqnarray}
c_1 & =  &  \frac{1}{2 \pi i} {\rm
Tr}(F-F') \nonumber\\
c_2 & =  &  \frac{1}{(2 \pi i)^2}  \left ({\rm Tr}(F\wedge F-F'\wedge F') -
({\rm Tr}F \wedge {\rm Tr}F -{\rm Tr}F' \wedge {\rm Tr}F')\right)\nonumber\\
c_3 & = &  \cdots
\end{eqnarray}

The generalized fluxes, $\int c_i$, are quantized and thus measure the
embedded D-brane charges in a $\dap$ system -- meaning the number of
branes left after tachyon condensation~\cite{witten,audio}.  For
example, $\int c_1$ measures the number of $p-2$ branes minus the
number of $p-2$ anti-branes, or rather the total $p-2$ brane charge
left after tachyon condensation; $\int c_2$ measures the total $p-4$
brane charge of the system; $\int c_3$ measures the total $p-6$ brane
charge, etc.

\section{{\bf Brane Production in the Early Universe}}\label{production}

We have learned that lower dimensional branes result from tachyon
condensation when magnetic fluxes are turned on.  The central result
of this paper is that the fluxes needed for brane formation can be
generated dynamically in an expanding universe by the Kibble
mechanism. This is the same mechanism which generates defects such as
cosmic strings, textures and monopoles from GUT phase transitions.
These defects also require fluxes to be turned on and the Kibble
mechanism provides them through cosmological dynamics.

At high temperature, $T> T_c$, the tachyon
potential will be parabolic and a system of $N$ $\da9$ branes will be
stable against decay.  The energy density in the brane tensions $ 2 N
\tau_9$, the matter density on the brane due to gauge fields, and
massive modes, will act as a source for the expansion of the $\da9$
branes. If the expansion is adiabatic, the temperature will decrease
monotonically.

\EPSFIGURE[l]{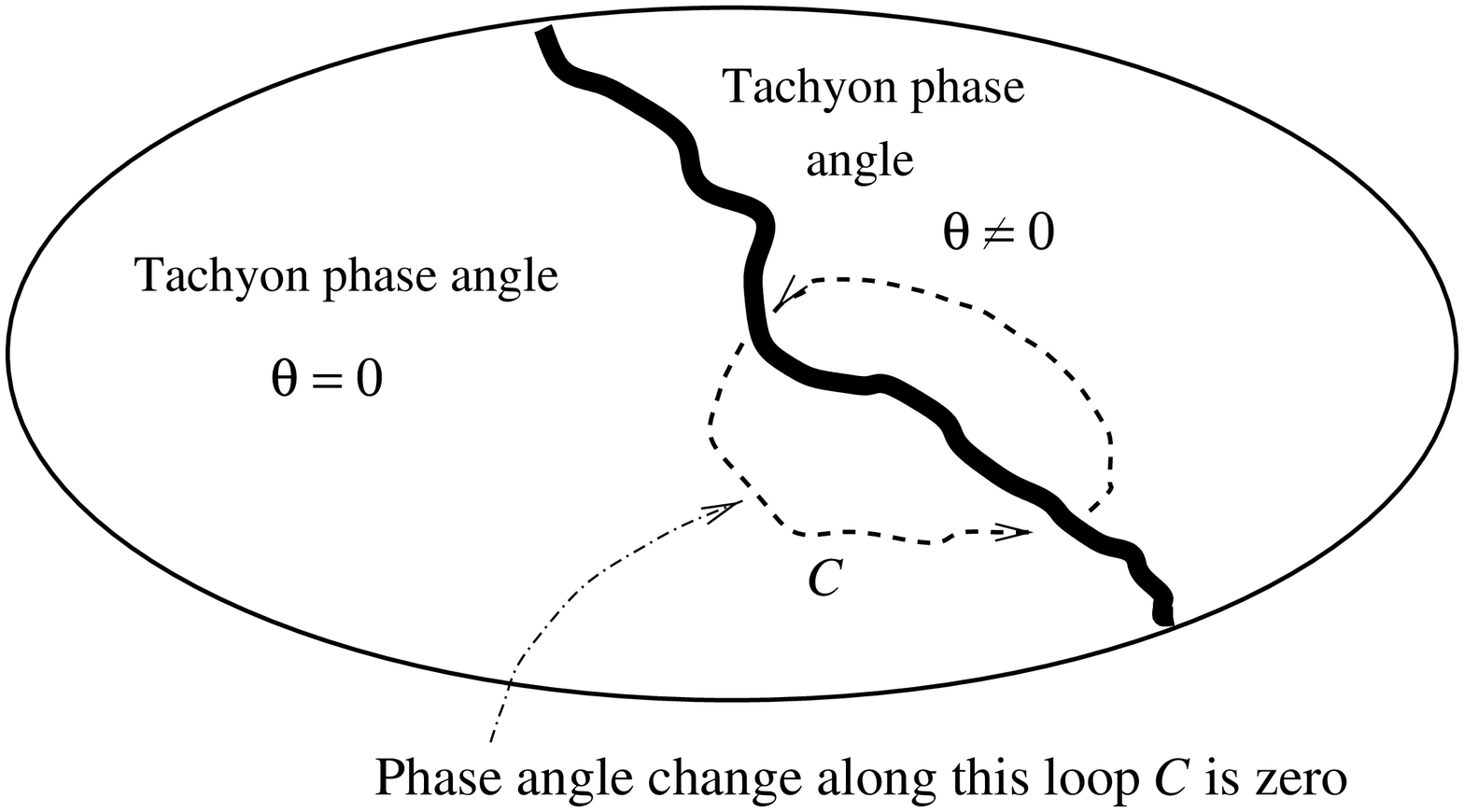,width=7cm}{8D domain wall in
$\IR^{10}$. The loop $C$ does not wind the vacuum manifold since
$\Delta \theta \neq 2 \pi$ around the loop. \label{domainwall}}

When the temperature approaches $T_c$, tachyonic instabilities will
arise as the tachyon potential takes on a double well shape.  After
the tachyon rolls down to the vacuum manifold $\cm$, it will be
characterized by a set of phase angles, $\{ \theta_i \}$, and a
modulus $|t_0|$. For example, if $\cm = S^1$, then the tachyon will
take the expectation value $\langle t \rangle =|t_0|e^{i\theta}$.
However, near the temperature $T_c$, the tachyon field will fluctuate
randomly, rolling down the potential and then rolling up via thermal
fluctuations.  Hence $t$ will not take on any definite values and the
phase angles will fluctuate.  Once the temperature falls below $T_c$
and reaches the Ginzburg temperature $T_G$, thermal fluctuations will
no longer be able to push the tachyon up the potential again.  At this
point, once the tachyon rolls to the bottom of $V(t)$, its phase will
be frozen in. Note, the Ginzburg temperature is close to $T_c$ and can
be written as $h(g_s) T_c$, where $h(g_s)$ depends only on the string
coupling and is typically close to unity~\cite{shellard}.

In a second order phase transition such as the tachyonic transition,
the correlation length of the field approaches infinity.  However,
expanding universes have causal horizons, which bound the distance
over which causal processes can occur.  In a universe with a Hubble
parameter $H\sim 1/t$, causal processes can occur only within a sphere
of diameter $H^{-1}$.  Thus an expanding universe will have regions
which are causally disconnected from each other.

Suppose that no topological obstructions in the usual sense appear
such that all $N$ $\da9$ pairs condense completely to vacuum.  Then
the tachyon field will have a magnitude $|t_0|$ everywhere.  However,
because Hubble volumes are causally disconnected and since the
tachyon's phase on $\cm$ is randomly determined, the tachyon's phase
will generally be different in different Hubble volumes.  Spacetime
will thus possess a domain type structure, with the expectation value
$\langle t \rangle$ varying from Hubble region to region in a
relatively random way.  The question answered by Kibble about
cosmological phase transitions (like our tachyonic transition) was
whether any residue of false vacuum remains anywhere.  In particular
can false vacuum be trapped at the intersection points of Hubble
regions like flux tubes are trapped in a superconductor~\cite{kibble}?

\EPSFIGURE[l]{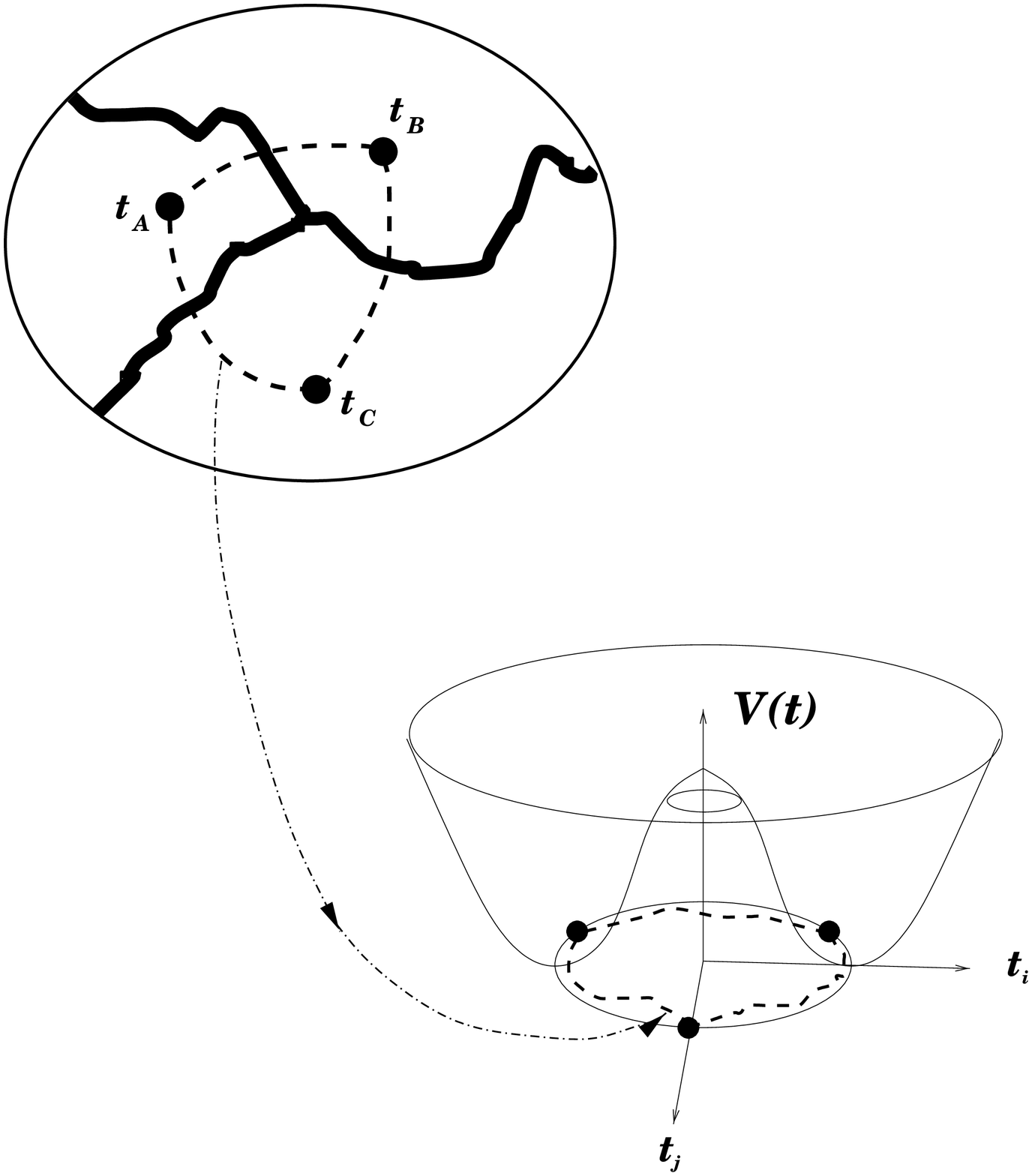,width=7cm}{Three Hubble volume sized
regions $A,B,C$, intersect at a point $P$, which is actually a 7D
edge.  The tachyon field takes the values $t_A, t_B,t_C$ on $A,B$ and
$C$ respectively.  The tachyon field maps the upper loop in spacetime,
$D$, which encloses $P$ to the lower loop on the vacuum manifold, $E$,
and is an element of $\pi_{1}(\cm)$. \label{homotopy}}

The answer depends on the shapes of the Hubble regions and how they
intersect.  Each region is nine-dimensional.  When two regions meet as
in figure \ref{domainwall}, their common boundary is 8D. There is a
jump of phase angles across the boundary.  However, the change in
phase angle(s) around the closed curve $C$ is not a non-zero multiple
of $2\pi$.  Thus no winding of the potential occurs.  However, if
three cells meet along an edge as in figure \ref{homotopy}, where the
edge is seven dimensional and coming out of the paper, then the phase
change around the closed curve $D$ enclosing the edge at $P$ will be
$2\pi$.  The tachyon maps $D$ onto the locus $E$ winding
$\cm$. Attempts to shrink the curve in spacetime will cause the path
$E$ to move off of $\cm$ and upwards to the false vacuum $t=0$.  Thus
along the edge, which is the intersection of the cells, a line defect
of false vacuum will be trapped.  This line defect is actually seven
dimensional and corresponds to a $D7$ brane.  The tachyon mapping the
spacetime circle, $D$, to the circle on $\cm$ is an element of
$\pi_1(\cm)=\IZ$.  The non-triviality of $\pi_1(\cm)$ can be
interpreted to mean that there exists a configuration, notably the 3
intersecting cells, for which a $S^1$ in spacetime can be mapped to a
locus winding the vacuum manifold. The winding insures that some false
vacuum exists, because to conserve winding number as the spacetime
circle is shrunk to zero size the tachyon must move off of $\cm$ to
the top of $V(t)$ where $t=0$.  Three cells were needed because three
points are needed to uniquely to determine a $S^1$.

Likewise, the non-triviality of $\pi_3(\cm)$ can be interpreted to
mean that there exists a configuration, notably five intersecting
cells, for which a $S^3$ in spacetime can be mapped to a locus winding
the vacuum manifold.  Five cells are  needed because five points
uniquely determine a $S^3$. (Think of the $S^3$ as passing through the
centers of five 4D balls, each of which represents a Hubble volume
with a different $\langle t \rangle$.) The intersection of the cells is a 5D
subspace. Winding then corresponds to the presence of a $D5$ brane
at the intersection point.

In general a sphere $S^k$ is determined by $k+2$ points\footnote{To
make this obvious use induction and the fact that a $S^1$ is
determined by an inscribed triangle (three points), and to add an
extra dimension to a polyhedron:  add a point in the extra
dimension. More pedanticly, $S^n$ is the locus $\sum_1^{n+1} (x_i -
\bar{x}_i)^2 = R^2$.  All the parameters
$(\bar{x}_1,\ldots,\bar{x}_{n+1},R)$ can be determined by $n+2$
equations, i.e. $n+2$ points.} If each point corresponds to a
different vacuum expectation value $\langle t_i \rangle$, then
$\pi_{k}(U(N)) =\IZ$ corresponds to $k+2$ cells intersecting in a
$9-(k+1)$ subspace with a $D(9-(k+1))$ or $\bar{D}(9-(k+1))$ brane
trapped at the intersection.  See Table \ref{table}.

\begin{table}
\begin{center}
\begin{tabular}{|c|c|c|c|c|c|c|c|c|c|c|} \hline
\ $Dp$-brane\ & \ $D7$\ & \ $D6$\ & \ $D5$\ & \ $D4$\ & \
$D3$\ & \ $D2$\ & \ $D1$\ & \ $D0$\ \\ \hline
Loop in $\IR^{10}$ &
$S^1$ & $S^2$ & $S^3$ & $S^4$ & $S^5$ & $S^6$ & $S^7$ & $S^8$   \\ \hline
No. regions & 3 & 4 & 5 & 6 & 7 & 8 & 9 & 10 \\ \hline
Type IIB charge &$\IZ$ & 0
& $\IZ$ & 0 & $\IZ$ & 0 & $\IZ$ & 0  \\ \hline
\end{tabular}
\end{center}
\caption{\em The number of Hubble regions with a common intersection
needed to produce a $Dp$ brane; the last line gives the possible Type
IIB $D$-brane charge at such an intersection.}
\label{table}
\end{table}

Thus because $\pi_7(U(N))=\pi_5(U(N))=\pi_3(U(N))=\pi_1(U(N)) = \IZ$
for sufficient $N$, tachyon condensation on a group of $\da9$ pairs
will generically create either $D7$ or $D5$ or $D3$ or $D1$ branes at
the intersections of different Hubble regions.  The specific $D$-brane
which forms will depend on the dynamics of the Hubble regions and how
they intersect.  Table \ref{table} shows that lower dimensional branes
require more regions to have a common intersection.  Although the
regions are irregularly shaped, it is hard to force many regions to
have a common boundary.  From this point of view, it would seem that
higher dimensional branes are more likely to be created.  However, as
we show later, energetics and entropy arguments favor lower
dimensional defects.

It is interesting that the spectrum of string theory defects which
cosmology creates via the Kibble mechanism is much richer than in
GUTs.  In order to form different defects like monopoles and
cosmic strings in GUTs,  one needs a complicated multi-step symmetry breaking
mechanism.  In our case, different defects appear because the
spacetime has more dimensions; thus more homotopy groups are relevant.
The $\pi_{2k-1}$ can be made  non-trivial simply by increasing $N$
which enlarges the unbroken gauge group $U(N) \times U(N)$ and the
broken symmetry group $U(N)$.

Since we can roughly associate one intersection to each Hubble volume,
a {\em lower bound} on the number of branes formed by tachyon
condensation in an expanding universe is: one brane per Hubble volume.

Below, we point out another mechanism by which the Kibble mechanism
can operate which may be more realistic.  However, it is based on
time-dependent aspects of tachyon condensation which are not well
understood.

The tachyon field is actually a matrix and consists of $N^2$
individual tachyon fields $t_{ij}$ where $0\le i,j\le N$.  It is
charged by the $U(N)$ gauge fields on the branes and the $U(N)$ gauge
fields on the anti-branes.  Thus, it transforms under the bifundamental
representation of $U(N)\times U(N)$, which is broken to $U(N)$ when
the tachyon field receives an expectation value on the vacuum manifold
$\cm$.  Using the gauge symmetry, one can diagonalize the tachyon $t$,
at every point in its evolution, and isolate the degrees of freedom in
$N$ diagonal tachyon fields $t_{ii}$ which we call $t_i$. We
label time by $\tau$. The $t_i$ are independent of one another.
However, in the diagonal basis their classical equations of motion in
flat space are the same:\footnote{$V'(t_i)$ rather than $V'(t)$,
because in the diagonal basis the effective potential can be written
as $V(t) = 2N\tau_9 - c_1g_s^{-1}(t_1^2 + \cdots + t_n^2) +
c_2g_s^{-2} (t_1^4 + \cdots + t_n^4) + {\cal{O}}(g_s^{-3})$} $ \Box
t_i = -V'(t_i)$, and the initial conditions are also the same
$t(\tau=0) =0$.  Thus classically at each point in time, the fields
$t_i$ should be the same, $t_1(\tau) = t_2(\tau)= \cdots = t_N(\tau)$.
However, at the temperature $T_c$, the fields will feel thermal
fluctuations, as well as quantum perturbations of magnitude
 $\Delta t_i = H/(2\pi)$.  Hence, as the tachyons roll up and
down the potentials $V(t_i)$ they will lose coherence and generally
$t_1(\tau) \neq t_2(\tau)\neq \cdots \neq t_N(\tau)$, since the
evolution of the tachyon field will be partially stochastic

\begin{equation}
dt_i = t_i(\tau)\mu(\tau) d\tau +  t_i(\tau) \sigma^2(\tau) dW_i.
\label{stochastic}
\end{equation}
\n
The first term in (\ref{stochastic}) is the deterministic
component of $t_i$'s evolution. The second piece is a Brownian
motion.

The lack of coherence of the $t_i$ will not break the gauge symmetry
of $t$.  However, the lack of coherence means that as the Ginzburg
temperature is approached the final values of the $t_i$ will be frozen
in at different times. This effectively means that the $N$ $\da9$
brane anti-brane pairs annihilate at different times.  For example,
suppose $t_3,\ldots,t_N$ undergo their final condensation at the same
time, and suppose that $t_1 = t_2$ and that $t_1$ and $t_2$ have their
final values frozen in several string times after $t_3,\ldots,
t_N$. Hence, while $t_1$ and $t_2$ are still oscillating, the other
tachyons will have disappeared and the brane anti-brane pairs and the
gauge fields on their worldvolume will have as well.  Thus, $t_1, t_2$
will no longer see the gauge fields on the $N-2$ branes which have
condensed (because the gauge fields will have become confined.)  They
will only see the gauge fields on the first and second brane pair and
will only feel a $U(2)$ gauge symmetry. Hence the gauge symmetry on
$\cm$ will have been broken to $U(N) \rightarrow U(N-2) \times U(2)$.
The last two branes can lead to brane production via the Kibble
mechanism.  In particular they can create 7-branes or 5-branes, since
$U(2)$ is in the stable range of only $\pi_1$ and $\pi_3$.  The first
$N-2$ pairs may typically produce lower dimensional branes, since
$U(N-2)$ is in the stable range of all the relevant homotopy groups,
$\pi_1,\pi_3,\pi_5,\pi_7,\pi_9$ if $N$ is large.

More generally, a group of $N$ tachyons will not all condense at the
same time because of perturbations and because the phases are not
frozen in exactly at $T_G$, but in a narrow band around $T_G$. If the
condensation is staggered such that $n_1$ tachyons first condense at
roughly the same time, and $n_2$ condense a few strings times later,
and so on until the last group of $n_q$ tachyons condense, then the
gauge symmetry on $\cm$ will be broken to $U(N) \rightarrow
U(n_1)\times U(n_2) \times \cdots \times U(n_q)$. The number of
possible ways to condense is equal to the number of subsets of $N$
brane pairs, or rather the power set of $N$, which is $2^N$. This is
an enormous number, and it is statistically likely that many groups
$n_j \sim {\cal{O}}(1)$ will exist.  Such $n_i$ can only lead to the
production of high dimensional branes since $U(n_i)$ is in the stable
range of only the lower homotopy groups.  If $n_j \gg 1$, then very
low dimensional branes may result from tachyon condensation and the
Kibble mechanism, as $U(n_j\gg 1)$ is in the stable range of all the
relevant homotopy groups.  This multi-step symmetry breaking is a
generalization of elaborate multi-step symmetry breaking patterns in
GUTs which form multiple defects like monopoles and cosmic strings.

The $q$ partitions of $n_i, 0 \le i \le q$, imply that a lower bound
on the number of branes which form is: $q$ $D$-branes per Hubble
volume.  These branes will typically intersect since they are extended
objects, and a network of branes analogous to a string network will
fill spacetime.

To determine whether the {\em underlying physics} prefers the creation of
large dimensional branes ($n_i \gg 1 $) or small dimensional branes
($n_i \sim 1$), we employ some crude arguments based on energetics and
entropy maximization.  From the point of view of energetics, one would
expect the tachyons to condense in as many dimensions as possible
unless the tachyon profile was set by hand.  Regions where the tachyons
have not condensed are filled with the false vacuum $t=0$, and possess
a higher energy $V(0)$, than regions where the tachyons have condensed
-- regions where $V = V(t_0)$.  From a different perspective, regions
which are filled with D-branes have a sizable vacuum energy since
D-branes are heavy at weak string coupling.  Alternatively, from an
entropy point of view one might expect the large possible entropy of
the gauge theory on the branes to make it entropically favorable to
have large regions of false vacuum.

%To calculate ${\rm dim}(\ct)$ we perform a very rudimentary
%calculation and estimate the relative probability for the creation of
%a $p$-brane as opposed to a $p'$-brane or nothing at all.  If ${\rm
%dim}(\ct)=p$ then $Dp$ branes may be produced, and if ${\rm
%dim}(\ct)=p'$ then $Dp'$ branes may be produced, etc. This calculation
%is difficult to do from the microscopic physics.  Thus, we use a
%simple scheme based on entropy maximization.

Consider an initial state $i$ of $N$ $\da9$ pairs filling a compact
toroidal spacetime. Let the toroid be square and have a radius of
$r_0$.  We would like to calculate the probability for the state $i$
to decay into a final state $f_p$ relative to the probability for $i$
to decay into a different state $f_{p'}$.  We will take the state
$f_p$ to consist of a gas of $Dp$ branes wrapping $p$ cycles of the
torus and a gas of gravitons in the transverse space.  The state
$f_{p'}$ is similar except $Dp'$ branes wrap $p'$ cycles and gravitons
fill the remaining transverse space.  For simplicity, we will assume
that the final state branes are not coincident and that no coincident
brane anti-brane pairs are left.

The probability of a  state, $i$, is proportional to the Euclidean
partition function

\begin{equation}
P(i) \sim Z(i)  \sim e^{-F(i)} = e^{-(U(i)-TS(i))}
\end{equation}

\n where $F(i)$, $U(i)$ and $S(i)$ are the free energy, energy and
entropy of the state $i$ respectively.

The relative probability for the decay process $i\rightarrow f_p$ is
then

\begin{equation}
{\cal{P}}(i\rightarrow f_p) \equiv \frac{P(f_p)}{P(i)} \sim \frac{e^{-F(f_p)}}{e^{-F(i)}} \sim
e^{T(S(f_p) -S(i))}
\label{prob}
\end{equation}

The entropy of the initial state, $S(i)$, is the sum of the entropy of the
tachyon and gauge field gas on the world volume of the $\da9$ branes
which is denoted $S_{t+g}$, and the entropy of any embedded solitonic
configurations, $S_{soliton}$. Thus

\begin{equation}
S(i) = S_{t+g} + S_{soliton}
\end{equation}

The relative probability for the two processes
 ${\cal{P}}(i\rightarrow f_p)$ and ${\cal{P}}(i\rightarrow f_p')$ is
 then, using Eq. \ref{prob},

\begin{equation}
\frac{{\cal{P}}(i\rightarrow f_p)}{{\cal{P}}(i\rightarrow f_{p'})}
\equiv \frac{P(f_p)/P(i)}{P(f_{p'})/P(i)} \sim \frac{e^{-F(f_p)}}{e^{-F(f_{p'})}} \sim
e^{T(S(f_p) -S(f_{p'}))}
\label{relprob}
\end{equation}

Unsurprisingly, the relative probability depends only on the
difference in entropies of the states $f_p$ and $f_{p'}$.  The
entropy $S(f_p)$ is a sum of the entropy of the photon gas on the
$Dp$ branes $S_{\gamma}$, the entropy due to the motion of a $Dp$
brane $S_{Dp}$, and the entropy of the graviton gas in the
transverse space $S_{g}$.

\begin{equation}
S(f_p) = S_{\gamma}+ S_{Dp}+ S_{g} \label{final}
\end{equation}

\n We will ignore $S_{Dp}$ because it is negligible compared to
the entropies of the relativistic graviton and photon gases,
$S_{g}$ and $S_{\gamma}$.

The equation of state for a gas in $d$ spatial dimensions is $
p_{ress} = \rho/d$.  Here $p_{ress}$ is the pressure, and $\rho$
is the energy density.  The gravitons in Eq. \ref{final} move in
$9$ spatial dimensions and the photons move in $p$ spatial
dimensions. Thus, ~\cite{kolb},

\begin{eqnarray}
S_{g} & = & \frac{(p_{ress}+\rho_g)}{T} V_{g}= \frac{10}{9} \frac{
\rho_g V_{g}}{T}
= \frac{10}{9} \frac{E_{g}}{T}\nonumber\\
S_{\gamma} & = & \frac{(p_{ress}+\rho_{\gamma})}{T} V_{p}=
\frac{p+1}{p} \frac{ \rho_{\gamma} V_{p}}{T} = \frac{p+1}{p}
\frac{E_{\gamma}}{T} \label{entropygas}
\end{eqnarray}

\n where $T$ is the temperature, $V_p$ is  the volume of a $p$
dimensional torus, and $\rho_g, \rho_{\gamma}$ are the energy
densities of the graviton and photon gases. $E_{g}$ and
$E_{\gamma}$ are the energies of the graviton and photon gases.
The volume $V_g$ is the nine-dimensional volume $V_9$ minus the
volume of the daughter brane $V_p$.

\begin{equation}
V_g = V_9-V_p
\end{equation}

Energy conservation requires

\begin{equation}
E = E_{g} + E_{\gamma} + E_{Dp}
\end{equation}

\n where $E_{Dp} = \tau_p V_p$. The entropy $S(f_p)$ is then

\begin{equation}
T S(f_p) = \frac{10}{9} E_g + \frac{p+1}{p} E_{\gamma} =
\frac{10}{9} E + \left (\frac{1}{p} -\frac{1}{9}\right )
E_{\gamma} -\frac{10}{9} E_{Dp}
\end{equation}

To calculate the relative probability of obtaining state $f_p$ or
$f_{p'}$ in Eq. \ref{relprob}, we can calculate $\partial
S(f_p)/\partial p$ instead of the difference, $S(f_p)-S(f_{p'})$.

\begin{equation}
\frac{ \partial TS(f_p)}{\partial p}  = \left [ \left (\frac{1}{p}
-\frac{1}{9}\right )\rho_{\gamma} -\frac{10}{9} \tau_p \right ]
\frac{ \partial V_p}{\partial p} < 0 \ \ \ \  {\rm for}  \
\rho_{\gamma}(T) \ll \tau_p \label{rubbishequation}
\end{equation}

\n If $ \tau_p > \rho_{\gamma}(T)$ then the inequality
(\ref{rubbishequation}) is true.  This will happen if the brane is
not highly excited when produced.  Also, if the temperature at
which the brane is produced is lower than the string scale then
the inquality is valid.  This may occur if cosmological expansion
on the branes redshifts the energy density on the parent branes
and anti-branes during tachyon condensation.  Equation
(\ref{rubbishequation}) implies that the entropy increases as the
dimension of the daughter $Dp$ branes decreases. This is not
surprising. One expects it to be entropically favorable to produce
a gas of gravitons rather than a heavy D-brane and its
world-volume photon gas.  The tension of the daughter D-brane
soaks up most of the energy leaving little to excite the photon
gas on it.  If the energy of the photon gas exceeds the energy of
the brane tension, then a thick halo of open strings covers the
brane and the equality in (\ref{rubbishequation}) is reversed.
This is understandable as this is the regime in which the argument
in section \ref{motive} favoring the production of high
dimensional branes applies.

Thus, for realistic conditions entropy arguments favor the
production of lower dimensional branes over the production of
higher dimensional branes.  This contrasts with our earlier
argument that low dimension branes are more difficult to create
via the Kibble mechanism.  We believe, that entropy arguments are
likely to prevail, although this must be explicitly checked.

>From eq. (\ref{relprob}) one can argue that lower dimensional
branes are exponentially more likely than higher dimensional ones.
>From a brane-world point of view this may be very attractive, as
it implies that the production of branes with $p>3$ is
exponentially suppressed relative to the production of $D3$
brane-worlds.  This also implies that few $D3$ branes will be
produced, and that mostly $D1$ branes and perhaps $D(-1)$ branes
(which are instantons) will form.  However, if $N$ is large the
probability for the formation of a $D3$ brane becomes
non-negligible and is given by a Poisson-like process.
\footnote{An alternate means of dynamically creating $D3$
braneworlds was communicated to us by Kogan and Abel.  They
suggested that the thermal production of high dimensional branes
will also create embedded $D3$ branes which annihilate last among
the embedded branes on the worldvolume of the larger dimensional
branes and are the largest dimension branes to survive. See also
~\cite{kogan}.}

D-strings may form on the $D3$ branes as embedded defects when a $D1$
and a $D3$, formed from two partitions,  overlap.  Such configurations
are well known~\cite{douglas, gregory, semenoff,topoeffects}.  The embedded
D-strings may cause cosmological problems and may induce gravitational
collapse of the brane universes.  Monopoles were similar disasters for
pre-inflation cosmology. Inflation at high energies is needed to
dilute the $D1$ branes.\cite{fernando,inflation}

We now relax the condition that no topological obstructions to tachyon
condensation other than the Kibble mechanism appear.  In the very
early universe, the open strings on the space-filling branes and
anti-branes are expected to be highly excited.  The $\da9$ branes will
be covered with a cloud of excited perturbative and non-perturbative
phenomena.  The excitations on the branes and anti-branes will be
broadly similar.  However, large fluctuations may occur, and
typically, the fluxes and non-perturbative excitations will be
somewhat different on the the branes and anti-branes. This is
consistent, for example, with a primordial magnetic field possibly
generated some time before the era of structure formation. The Chern
classes, $c_m$ measure how the excitations differ on the branes
and anti-branes.  Unless the gauge field configurations on the branes
and anti-branes are topologically identical, the Chern classes will be
non-zero.  Hence, $N$ $\da9$ branes near the big bang will generally
possess embedded charges.

Represent the spacetime $M$ as a product manifold $M = X \times Y$,
where $Y$ is some compact submanifold.  Now

\begin{equation}
M = \sum_{\{x~\in X\}} X \times Y_x
\end{equation}

\n where the sum is over all points $x$ in $X$.  We think of $X$ as
fibered at every point $x$ by $Y$, where $Y_x$ is the fiber at $x$. If
$c_m$ integrated over $Y_x$ is non-zero, then after tachyon
condensation a $D(9-2m)$ brane will be left wrapping $Y$ at the
position $x$ on $X$.  Thus one $D(9-2m)$ brane will exist in one
Hubble volume surrounding $x$.  Unlike in the Kibble mechanism, there
is a brane in only one Hubble volume.  It is likely that $c_m$
integrated over $Y_x$, for many points $x$, will be non-zero.
However, the number of defects created in this way -- which relies on
random random cosmological excitations to impose $\int_{Y_x} c_m \neq
0$ -- can never compare to the number of defects created by the Kibble
mechanism.  However, if one so desires, one can turn on various fluxes
and carefully set the tachyon field, such that when the universe exits
from the tachyonic phase transition, the configuration of choice (a
susy intersecting brane configuration, etc.) will exist in at least
one place in the universe.

\section{Type IIA Theory}\label{typeIIAtheory}

\n All of our Type IIB results have analogs in Type IIA theory.
Instead of starting with $N$ $\da9$ brane anti-brane pairs as an
initial state, one starts with $2N$ non-BPS $D9$ branes (or $\bar{D}9$
branes) in Type IIA theory.  These branes are stable at high
temperature and become tachyonic at low temperature. The tachyon
potential has a double well structure, and if $2N$ $\a 9$ branes are
initially coincident, they will possess a $U(2N)$ gauge symmetry,
which is broken on the vacuum manifold to

\begin{equation}
\cm_{IIA} \sim \frac{U(2N)}{U(N)\times U(N)}
\end{equation}

\n because on $\cm$ the tachyon field possesses $N$ identical
eigenvalues $t_0$ and $N$ identical eigenvalues $-t_0$.

Topological obstructions to tachyon condensation may exist in the form
of lower dimensional brane charges.  However, the brane charges are measured
by the K-theory group ${\rm K}^1$, instead of ${\rm K}^0$ as in
Type IIB theory.  ${\rm K}^1$ maps to the homotopy groups

\begin{equation}
{\rm K}^1(S^{2k+1+r}) =  \left\{
\begin{array}{cll}
\IZ     & {\rm if} &r=0\\
0   & {\rm if} &r=1
\end{array}\right\} = \pi_{2k+r} \left (\frac{U(2N)}{U(N) \times U(N)}\right)\ \ \ \
\forall N > 2k+1+r ,~ k \in \IZ^+.\
\label{typeIIA}
\end{equation}

\n This time if the tachyon condenses in odd dimensions ($r=0$),
stable $8-2k$ branes may form if $N \ge 2k+1$.  If the tachyon
condenses in even dimensions ($r=1$), no stable branes will form.  As
before the the stable range, $N> 2k+1+r$, selects the physical states.

The K group can be mapped into cohomology, and the total Chern class
can be decomposed into Chern classes, each of whose integrals measure
the embedded $D$-brane charges.  For example, $\int c_1$ measures the
net $D8$ brane charge, etc.

However, in reality ${\rm K}^1$ doesn't always accurately measure Type
IIA brane charge.  For more general spacetimes, the stable range for
${\rm K}^1$ begins at $N=\infty$, and thus an infinite number of
branes are required to create lower dimensional
defects~\cite{witteninfinity}.  It turns out that the K-homology group
${\rm K}_1$ measures Type IIA D-brane charge for finite $N$
and general spacetimes.  ${\rm K}_1$ is dual to ${\rm K}^1$ in the
same way that homology is Poincare dual to cohomology.  By substituting
${\rm K}_1$ for ${\rm K}^1$, we find a stable range beginning at finite
$N$, and everything is thought to work out~\cite{khomology}.

The cosmological dynamics in Type IIA are similar to the dynamics in
Type IIB.  The matter on the $\bar{D}9$ branes causes adiabatic
expansion, the temperature drops, and then tachyons appear.  The
tachyons condense completely to vacuum everywhere, unless embedded
charges exist.

Even if no embedded charges exist, branes still appear via the Kibble
mechanism at the intersections points of causally disconnected Hubble
regions. The specific branes which form depend on the dynamics of how
the Hubble volumes intersect. As before, to produce a $D(8-k)$ brane
trapped at an intersection, $k+2$ Hubble regions with different
$\langle t \rangle$ need to intersect. Since $\pi_{even} (\cm_{IIA}) =
\IZ$, only even (spatial) dimensional branes may form at the
intersection. Entropy arguments suggest that lower dimensional branes
are preferred.  Thus a Type IIA universe emerging from a tachyonic
phase transition may be populated with many, many $D0$ branes, which
are essentially monopoles.

\section{Conclusions} \label{conclusions}

We have showed that tachyon condensation in an expanding universe
always produces $D$-branes as topological defects.  A lower bound on
the number of branes produced is one D-brane per Hubble volume.  The
dimensionality and shape of higher dimensional branes depend on the
tachyon field and its dynamics.  Entropically, lower dimensional
branes are exponentially favored over say, $D5$ or $D7$ branes. The
probability to create a $D3$ brane-world increases as $N$ increases.
This may be a viable way to dynamically create a brane-universe.  The
branes which are produced will generally be intersecting, and a ``brane
network'' will fill spacetime. Specific brane configurations such as
those used to construct brane-world models with realistic gauge
groups, or those used to trap gravity, may be produced by carefully
adjusting the tachyon profile.

However, our arguments may seem to suffer from the following flaws:

$\bullet$ We considered an initial state of $N$ $\da9$ branes, which
may seem unnatural.  But, in order to obtain gauge fields and gravity
in the early universe branes must be included.  The branes which would
most naturally serve to begin/continue the evolution of the universe
are space-filling branes, since they do not artificially label
directions as Dirichlet or Neumann.  More importantly, all lower
dimensional D-branes can be built from space-filling D-branes.  In
this sense space-filling branes are fundamental objects in string
theory.  In fact, the presence of space-filling branes is not exotic
at all.  For example, had we started with Type I theory, the only
other string theory naturally containing open strings and branes, 32
$D9$ branes would have then been required for anomaly
cancellation~\cite{polchinski}.  Thus an important moral may be: the
desire for gauge field excitations in the early universe makes
space-filling branes inevitable.  Moreover, if one chooses to stay in
Type II theory, once space-filling branes are included, coincident
space-filling anti-branes must also be included for anomaly
cancellation.  The inclusion of branes and anti-branes makes a
tachyonic phase transition like the one we have described inevitable.

$\bullet$ We have largely ignored compactification issues.  However,
one of the important features of the tachyon potential is its
background independence.  Thus, the process of tachyon condensation in
our model is independent of whether the branes are wrapped or curved,
etc.  However, the notion of a Hubble volume does depend on the
geometry.  If space is flat, then the Hubble length, $H^{-1}$, is
infinite, and the Kibble mechanism doesn't operate. Typically then,
one $Dp$ brane per {\em universe} may be produced (if any are).  If
branes are wrapped on compact cycles and the cycles expand, then the
Kibble mechanism operates while the cycles are expanding.  However, as
the expansion increases, much of the energy of expansion is
transferred to strings winding around the compact directions, which
may stop the expansion.  If tachyon condensation occurs when the
cycles have stopped expanding, then no branes are generically produced
as the Hubble length in this case is as large as the space.  Hence,
the scale at which the phase transition occurs is critical.  In more
well known theories producing defects like cosmic strings, the
properties of the defects also crucially depend on the energy scale of
the phase transition.

Tachyonic phase transitions are stringy phenomena and are expected to
occur near the string scale.  Strings winding compact directions
dominate the energy when the size of the system is much larger than
the string scale.  When $R \sim \l_s$, energy is partitioned equally
between open string winding modes and open string momentum modes since
$R\sim l_s$ is the self-dual radius of simple compactifications like
toroidal compactifications.  Thus, during a tachyonic phase
transition, winding modes do not dominate~\cite{brandenberger}.  We
therefore believe our results are largely independent of wrappings and
geometry.

However, the spectrum of possible D-branes which can form via tachyon
condensation on a general spacetime $M$ with various compact or
non-compact cycles will depend on the topology. The Type IIB spectrum
is determined by the K-theory group $K^0(\bar{M})$, where $\bar{M}$ is
a suitable compactification of $M$.  In general, $K^0(\bar{M}) \neq
K^0(\bar{\IR}^n)$, and the possible $D$-branes which can form in the
two spaces will differ.  For example, if a spacetime possesses some
non-supersymmetric cycles $Y$, then in the absence of any
stabilization mechanisms, tachyon condensation on $M$ will can never
leave stable $D$-branes wrapped on $Y$. Also, the stable range may
vary between different spacetimes. Analogous concerns apply to Type
IIA theory.  This does not detract from our analysis.  However, at the
onset of estimating the number of defects arising from tachyon
condensation on an initial state of say $N$ $\da9$ branes, one must
identify the spectrum of $D$-branes allowed on $M$.

$\bullet$ Our mechanism involves some non-perturbative physics.  For
example, large 't Hooft coupling $g_s N$ is required to stabilize the
tachyon field.  Also, there is some confusion regarding the open
string vacuum which flows to the closed string vacuum via tachyon
condensation.  Arguments derived from the spacetime action seem to
indicate that the open string vacuum has a strong effective coupling
$\sim V(t=0)^{-1}$ ~\cite{sen3,strongcoupling}. However, arguments
originating from the world sheet expansion suggest that it is weakly
coupled, since the expansion is an expansion in ``holes'' on the
world-sheet.  Holes are weighted by $\sim V(t=0)$, which is very
small~\cite{weakcoupling}.  But, even if the physics is
non-perturbative, at least the low temperature description is
apparently under control as evidenced by the explicit calculation of
the exact tachyon potential up to higher derivative
terms~\cite{gerasimov}.  In any case, non-perturbative effects are
nothing new with respect to topological defects.  Their formation is,
after all, inherently non-perturbative.

The arguments put forth in this paper also apply to scenarios in which
brane and anti-brane pairs collide and produce $D3$ brane
worlds.\cite{fernando} For example, suppose that a single parallel
$D5-\bar{D}5$ pair collides. Suppose that the branes are so large that
they contain many Hubble volumes\footnote{Such large branes may not be
stable.}.  Then once the branes come within $l_s$ of each other, open
string tachyons will form~\cite{susskind}, and the tachyons will
eventually condense to different values in different Hubble regions.
Thus one brane (possibly a $D3$) per Hubble volume will be created by
the Kibble mechanism.

We have succeeded in creating D-branes cosmologically. But, an
over-abundance is produced.  One defect per Hubble volume is enough to
over-close the universe~\cite{kolb}.  Somehow the branes must be
diluted or a brane anti-brane annihilation mechanism must set in
 decreasing their numbers. The branes and anti-branes produced by the
Kibble mechanism will feel a long range attractive potential, $V(r)
\sim -1/r^{7-p}$~\cite{lifschitz}.  This force presumably leads to
annihilation of some fraction of the branes in each Hubble volume.  In
a forthcoming article ~\cite{paper4} these, and associated issues are
addressed.

\acknowledgments{We thank Michael Green, Fernando Quevedo, Mohammad
Akbar, Tibra Ali, Tathagata Dasgupta, Sakura Schafer-Nameki,
M. Amir. I Khan, for discussions; Richard Szabo for email
correspondence; and C.B. Thomas for highlighting a useful
reference. We are grateful to Venkata Nemani, Phillipe Brax, Ian Kogan
and Carsten van der Bruck for careful readings of the manuscript at
various stages.  A.C.D thanks PPARC. M.M thanks the Isaac Newton
Trust, the Cambridge Commonwealth Trust, Hughes Hall, and DAMTP for
financial assistance.}

%%%%%%%%%%%%%%%%%%%%%%%%%%%%%%%%%%%%%%%%%%%%%%%%%%%%%%%%%%%%%%%%%%%%%%%%%%%%
%                      REFERENCES                            %
%%%%%%%%%%%%%%%%%%%%%%%%%%%%%%%%%%%%%%%%%%%%%%%%%%%%%%%%%%%%%%%%%%%%%%%%%%%%

\end{document}